\documentclass[aps,prd,preprint,groupedaddress,nofootinbib,showpacs]{revtex4}

\usepackage{graphicx,epsf,color,amsmath}
\usepackage[caption=false]{subfig}
\newif\ifdraft
\drafttrue
\newif\ifpreprint
\preprinttrue

\def\fig#1{Fig.~{\ref{#1}}}
\def\Fig#1{Fig.~{\ref{#1}}}

\def\sect#1{Section~{\ref{#1}}}

\def\spa#1.#2{\left\langle#1\,#2\right\rangle}
\def\spb#1.#2{\left[#1\,#2\right]}
\def\tree{{\rm tree}}

\def\lr{\leftrightarrow}

\def\tr{\, {\rm tr}}
\def\eps{\epsilon}
\def\e{\epsilon}

\def\M{{\cal M}}

\def\T{{\cal T}}

\def\nt{{n_3}}
\def\ns{{n_0}}
\def\na{{n_2}}
\def\nhf{{n_{1/2}}}
\def\nthf{{n_{3/2}}}

\def\eqn#1{Eq.~(\ref{#1})}
\def\Eqn#1{Equation~(\ref{#1})}
\def\eqns#1#2{Eqs.~(\ref{#1}) and~(\ref{#2})}

\def\N{{\cal N}}

\def\be{\begin{equation}}
\def\ee{\end{equation}}
\def\bea{\begin{eqnarray}}
\def\eea{\end{eqnarray}}
\def\ba{\begin{eqnarray}}
\def\ea{\end{eqnarray}}

\def\tree{{\rm tree}}
\def\oneloop{{\rm 1\hbox{-}loop}}
\def\twoloop{{\rm 2\hbox{-}loop}}

\newbox\charbox
\newbox\slabox
\def\s#1{{      
\setbox\charbox=\hbox{$#1$}
\setbox\slabox=\hbox{$/$}
\dimen\charbox=\ht\slabox
\advance\dimen\charbox by -\dp\slabox
\advance\dimen\charbox by -\ht\charbox
\advance\dimen\charbox by \dp\charbox
\divide\dimen\charbox by 2
\raise-\dimen\charbox\hbox to \wd\charbox{\hss/\hss}
\llap{$#1$} }}

\def\tree{{\rm tree}}

\def\LGB{{\cal L}_{\rm GB}}

\def\LRRR{{\cal L}_{R^3}}

\begin{document}

\ifpreprint
\hbox{\hskip0.3cm UCLA/16/TEP/103 \hskip8.6 cm SLAC--PUB--16905}
\fi

\vskip0.3cm

\title{Two-Loop Renormalization of Quantum Gravity Simplified}

\author{Zvi~Bern${}^{a}$, Huan-Hang Chi${}^b$, Lance Dixon${}^{b}$ and Alex Edison${}^a$}
\affiliation{
$\null$\\
${}^a$Mani L. Bhaumik Institute for Theoretical Physics\\
Department of Physics and Astronomy\\
University of California at Los Angeles\\
Los Angeles, CA 90095, USA \\
$\null$\\
${}^b$SLAC National Accelerator Laboratory\\
Stanford University\\
Stanford, CA 94309, USA\\
}

\begin{abstract}

The coefficient of the dimensionally regularized two-loop $R^3$
divergence of (nonsupersymmetric) gravity theories has recently been
shown to change when non-dynamical three forms are added to the
theory, or when a pseudo-scalar is replaced by the anti-symmetric
two-form field to which it is dual.  This phenomenon involves
evanescent operators, whose matrix elements vanish in four dimensions,
including the Gauss-Bonnet operator which is also connected to the
trace anomaly.  On the other hand, these effects appear to have no
physical consequences in renormalized scattering processes.  In
particular, the dependence of the two-loop four-graviton scattering
amplitude on the renormalization scale is simple.  In this paper, we
explain this result for any minimally-coupled massless gravity theory
with renormalizable matter interactions by using unitarity cuts in
four dimensions and never invoking evanescent operators.

\end{abstract}

\pacs{04.60.-m, 11.15.Bt \hspace{1cm}}

\maketitle

\section{Introduction}

Recent results show that the ultraviolet structure of gravity is much
more interesting and subtle than might be anticipated from standard
considerations.  One example of a new ultraviolet surprise is the
recent identification of ``enhanced ultraviolet cancellations'' in
certain supergravity theories~\cite{N4gravThreeLoops,N5FourLoop},
which are as yet unexplained by standard
symmetries~\cite{KellyAttempt}.  Another recent example is the lack of any
simple link between the coefficient of the dimensionally-regularized
two-loop $R^3$ ultraviolet divergence of pure Einstein
gravity~\cite{GoroffSagnotti,vandeVen} and the renormalization-scale
dependence of the renormalized theory~\cite{PreviousPaper}.  While the
value of the divergence is altered by a Hodge duality transformation that
maps anti-symmetric tensor fields into scalars,
the renormalization-scale dependence is unchanged.
In contrast, for the textbook case of
gauge theory at one loop the divergence and the renormalization-scale
dependence---the beta function---are intimately linked.  In
Ref.~\cite{PreviousPaper}, a simple formula for the
renormalization-scale dependence of quantum gravity at two loops was
found to hold in a wide variety of gravity theories.  In this paper we
explain this formula via unitarity.

As established by the seminal work of 't~Hooft and
Veltman~\cite{HooftVeltman}, pure gravity has no ultraviolet
divergence at one loop.  This result follows from simple counterterm
considerations: after accounting for field redefinitions, the only
independent potential counterterm is equivalent to the Gauss-Bonnet
curvature-squared term.  However, in four dimensions this term is a
total derivative and integrates to zero for a topologically trivial
background, so no viable counterterm remains.  Hence pure
graviton amplitudes are one-loop finite.  Amplitudes with four or more
external matter fields are, however, generally divergent.

At two loops pure gravity does diverge, as demonstrated by Goroff and
Sagnotti~\cite{GoroffSagnotti} and confirmed by van de
Ven~\cite{vandeVen}.  The pure-gravity counterterm, denoted by $R^3$, 
is cubic in the Riemann curvature.
The two-loop divergence was recently reaffirmed
in pure gravity~\cite{PreviousPaper}, and was also studied in a variety of
other theories, by evaluating the amplitude for
four identical-helicity gravitons.  The actual value of the
dimensionally-regularized $R^3$ divergence changes when three-forms are 
added to the theory, even though they are not dynamical in four space-time
dimensions.  More generally, when matter is incorporated into the
theory, the coefficient of the $R^3$ divergence
changes under a Hodge duality transformation.  However, 
such transformations appear to have no physical consequences for 
renormalized amplitudes~\cite{PreviousPaper}.

The dependence of the two-loop divergence on duality transformations
is closely connected to the well-known similar dependence of the
one-loop trace anomaly~\cite{DuffInequivalence}.  One-loop 
subdivergences in the computation include those
dictated by the Gauss-Bonnet term, whose
coefficient is the trace anomaly~\cite{PreviousPaper}.  Duff and van
Nieuwenhuizen showed that the trace anomaly changes under duality
transformations of $p$-form fields, suggesting that theories related
through such transformations might be quantum-mechanically
inequivalent~\cite{DuffInequivalence}. Others have argued that these
effects are gauge artifacts~\cite{OneLoopEquivalence,
  SiegelEquivalence, FradkinTseytlinEquivalence}.  For graviton
scattering at two loops in dimensional regularization, quantum
equivalence can be restored, but only after combining the bare
amplitude and counterterm contributions~\cite{PreviousPaper}.

The surprising dependence of the two-loop $R^3$ divergence in gravity
on choices of field content outside of four dimensions emphasizes the
importance of focusing on the renormalization-scale dependence of
renormalized amplitudes as the proper robust quantity for
understanding the ultraviolet properties.  The divergence itself, of
course, never directly affects physical quantities since it can be
absorbed into a counterterm.  In contrast, the renormalization scale
dependence does affect physical quantities because it controls
logarithmic parts of the scaling behavior of the theory.  While this
is well known, what is surprising is that, in contrast to gauge
theory, the two-loop divergences of pure gravity are not linked in any
straightforward way to the scaling behavior of the theory.  An underlying
cause is that evanescent operators, such as the Gauss-Bonnet term,
contribute to the leading two-loop $R^3$
divergence of graviton amplitudes~\cite{PreviousPaper}.

Evanescent operators are well-studied in
gauge theory (see e.g.~Ref.~\cite{Evanescent}), where they can modify
subleading corrections to anomalous dimensions or beta functions.  A
standard one-loop subdivergence is associated with the one-loop matrix
element of a non-evanescent operator; integrating over the
remaining loop momentum generates a double pole $1/\e^2$ in the dimensional
regulator $\e=(4-D)/2$.  When the operator is evanescent, the matrix
element is suppressed in the four-dimensional limit, typically
reducing the double pole to a simple pole, but still leaving a
contribution to the anomalous dimension.  A key property that is
special to the two-loop gravity computation is that the divergent
evanescent contribution
begins at the same order as the first divergence.  However, similar
effects could appear in other contexts.  For example, in the effective
field theory of flux tubes with a large length $L$, there is an
evanescent operator which would otherwise contribute to the energy at
order $1/L^5$~\cite{OferZohar}; presumably it will have to be taken
into account in a dimensionally-regularized computation of
$(\ln L)/L^7$ corrections to the energy.

In contrast to the divergence, the renormalization-scale dependence
does appear to be robust and unaltered by duality transformations or
other changes in regularization scheme.  Indeed, a simple
formula was proposed~\cite{PreviousPaper} for
the $R^3$ contribution to this dependence at two loops,
which is proportional to the number of four-dimensional bosonic minus 
fermionic degrees of freedom.
Yet in Ref.~\cite{PreviousPaper} this simple formula only arose
after combining the dimensionally-regularized two-loop amplitude with
multiple counterterm contributions.  Intermediate steps involved
evanescent operators and separate contributions did not respect
Hodge duality; nor would they have respected supersymmetry if
we had treated fermionic contributions in the same way.

The purpose of the present paper is to explain the simple
renormalization-scale dependence in terms of unitarity cuts
in four dimensions.  This approach turns a two-loop computation
effectively into a one-loop one, it manifestly respects Hodge duality 
and supersymmetry, and evanescent operators never appear.

This paper is organized as follows:  In \sect{ReviewSection} we summarize
the previous approach of Ref.~\cite{PreviousPaper}, along with the
the surprisingly simple formula found for the renormalization-scale dependence 
of the four-graviton amplitude at two loops.  Then in \sect{FourDimSection}
we derive the formula purely from four-dimensional unitarity cuts.
Our conclusions are given in \sect{ConclusionSection}.


\section{Review of previous approach}
\label{ReviewSection}

Pure gravity is described by the Einstein-Hilbert Lagrangian,
\begin{align}
{\cal L}_{\rm EH} & = - \frac{2}{\kappa^2} \sqrt{-g} R\,,
\label{Lagrangian}
\end{align}
where $\kappa^2 = 32 \pi G_N = 32 \pi/M_P^2$ and the metric signature
is $({+}{-}{-}{-})$.  While we are primarily interested in pure gravity,
it is insightful to include matter as well, as in Ref.~\cite{PreviousPaper}, 
by coupling gravity to $\ns$ scalars, $\na$ two-forms and $\nt$ three-forms,
as well as fermionic fields, $\nhf$ of spin-$1/2$ and $\nthf$ of
spin-$3/2$.

At one loop, graviton amplitudes do not diverge in four dimensions, because
no viable counterterms are available after accounting for field
redefinitions and the Gauss--Bonnet (GB) theorem~\cite{HooftVeltman}.
Divergences do occur if we allow the fields to live outside of four
dimensions~\cite{ConformalAnomaly,GoroffSagnotti,DuffInequivalence}.
The Gauss--Bonnet counterterm is given by
\begin{align}
\LGB\ =&\ \frac{1}{(4 \pi)^2} \frac{1}{\eps}
\Bigl( \frac{53}{90}+ \frac{\ns}{360} + \frac{91 \na}{360}
- \frac{\nt}{2} 
+ \frac{7 \nhf}{1440} - \frac{233 \nthf}{1440} \Bigr) \times \sqrt{-g} 
(R^2 - 4 R_{\mu\nu}^2 + R_{\mu\nu\rho\sigma}^2) \,.
\label{GaussBonnetDivergence}
\end{align}
At one loop, matter self-interactions cannot affect
this graviton counterterm.  The divergence represented by
\eqn{GaussBonnetDivergence} vanishes for 
any one-loop amplitude with four-dimensional external gravitons.
Amplitudes with four external matter states generically have divergences
in four dimensions, starting at one loop.
We neglect such divergences in this paper because they
do not affect the two-loop four-graviton divergence.

In the context of dimensional regularization, evanescent operators,
whose matrix elements vanish in four dimensions, can contribute
to higher-loop divergences.  Indeed, the Gauss--Bonnet term generates
subdivergences at two loops, because the momenta
and polarizations of internal lines can lie outside
of four dimensions~\cite{Capper,PreviousPaper}.

The coefficient in front of \eqn{GaussBonnetDivergence} has a rather
interesting story, because it is proportional to the trace
anomaly~\cite{ConformalAnomaly,GoroffSagnotti,DuffInequivalence}.
The connection comes about because the
calculations of the ultraviolet divergence and the trace anomaly are
essentially identical, except that in the latter calculation 
we replace one of the four graviton polarization
tensors with a trace over indices.  As already noted,
the trace anomaly has long been
known to have the rather curious feature that it is not
invariant under duality transformations~\cite{DuffInequivalence} that
relate two classical theories in four dimensions.  
In more detail, under a Hodge duality
transformation, in four dimensions the two-form field is equivalent to
a scalar and the three-form field is equivalent to a cosmological-constant
contribution:
\be
H_{\mu\nu\rho}\ \lr\ \frac{i}{\sqrt{2}} \,
\varepsilon_{\mu\nu\rho\alpha} \, \partial^{\alpha}\phi \,,
\hskip 2.5 cm 
H_{\mu\nu\rho\sigma}\ \lr\
\frac{2}{\sqrt{3}} \, \varepsilon_{\mu\nu\rho\sigma} \, 
 \frac{\sqrt{\Lambda}}{\kappa} \,.
\ee
\Eqn{GaussBonnetDivergence} shows that the trace anomaly, and hence
the associated evanescent divergence, change under duality transformations:
The coefficients in front of $n_2$ and $n_0$ differ, and the one in
front of $n_3$ is nonzero.
Correspondingly, subdivergences in two-loop amplitudes depend on
the field representation used.

In contrast to one loop, at two loops pure gravity in four dimensions
does diverge in dimensional regularization,
as shown by Goroff and Sagnotti~\cite{GoroffSagnotti} and
confirmed by van de Ven~\cite{vandeVen}.
In the $\overline{\rm MS}$ scheme, with $\e=(4-D)/2$,
the divergence is given by
\be
\LRRR^{\rm div}\ =\ 
-\frac{209}{1440} \left(\frac{\kappa}{2}\right)^2
\frac{1}{(4 \pi)^4} \frac{1}{\eps} 
\sqrt{-g}\, R^{\alpha \beta}{\!}_{\gamma\delta} 
R^{\gamma \delta}{\!}_{\rho\sigma} R^{\rho\sigma}{\!}_{\alpha\beta} \,.
\label{GSDiv}
\ee
In this computation, a mass regulator was introduced,
in addition to the dimensional regulator, in order to deal with certain
infrared singularities.  This procedure introduces regulator dependence
which is removed by subtracting subdivergences, integral by integral.
The subdivergence subtraction also properly removes
the Gauss--Bonnet subdivergences, leaving only the two-loop 
divergence.

\begin{figure}[tb]
\begin{center}
\captionsetup[subfigure]{labelformat=empty} 
\subfloat[\large (a)]{\includegraphics[scale=.38]{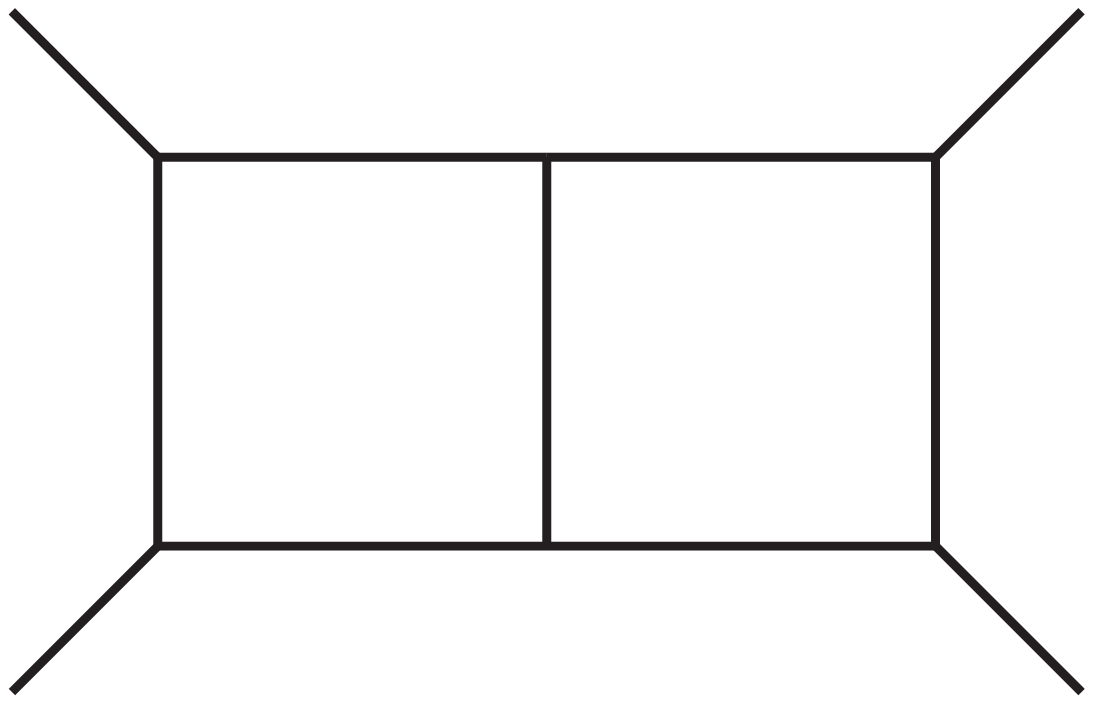}}
\hspace{.3cm}
\subfloat[\large (b)]{\includegraphics[scale=.38]{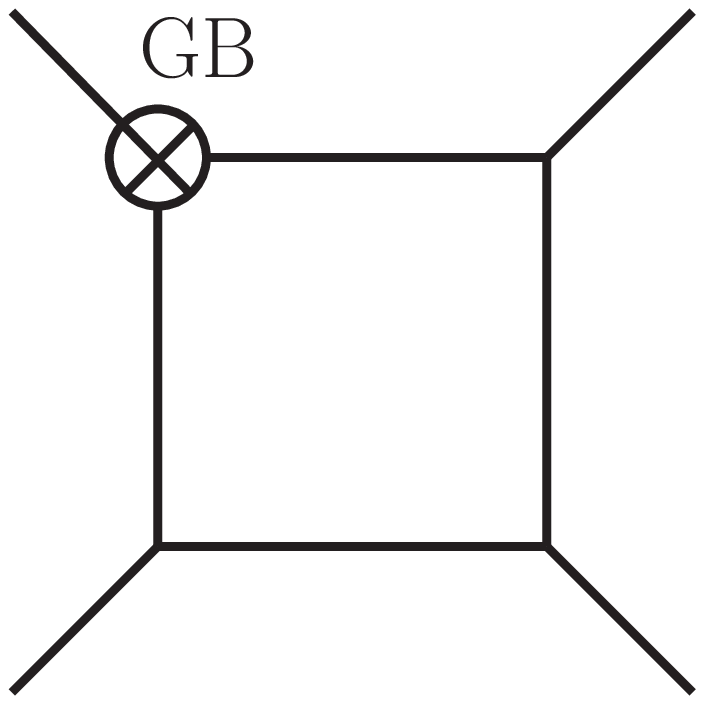}}
\hspace{.3cm}
\subfloat[\large (c)]{\raisebox{.55cm}{\includegraphics[scale=.38]{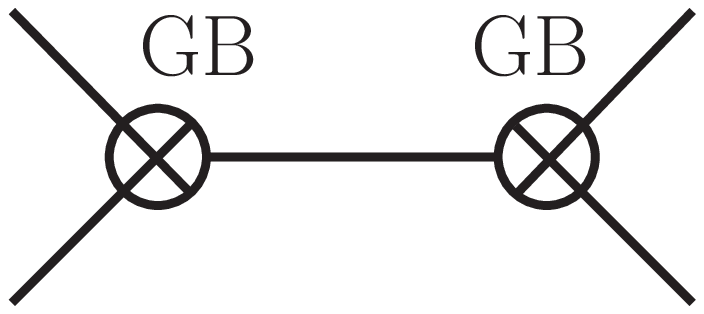}}}
\hspace{.3cm}
\subfloat[\large (d)]{\raisebox{.15cm}{\includegraphics[scale=.38]{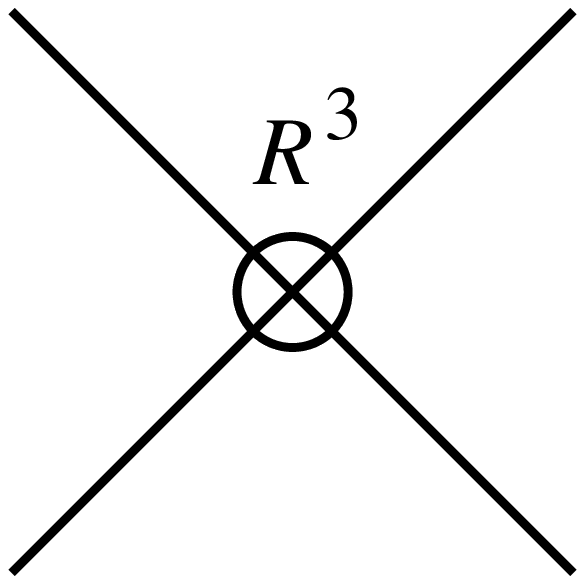}}}
\end{center}
\vskip -.7cm 
\caption[a]{\small Representative four-point diagrams for
  (a) the bare contribution, and the
  (b) single-GB-counterterm, (c) double-GB-counterterm, and
  (d) $R^3$-counterterm insertions needed to remove all divergences.
\label{CounterSingleDoubleFigure}
}
\end{figure}

In Ref.~\cite{PreviousPaper}, the same $R^3$ divergence (\ref{GSDiv})
was extracted from a four-graviton scattering amplitude with all
helicities positive, $M_4^{\rm 2-loop}({+}{+}{+}{+})$.
This helicity amplitude is particularly simple to calculate,
making it a useful probe of the two-loop ultraviolet structure.
It is sensitive to the $R^3$ operator because the
insertion of $R^3$ into the tree amplitude gives a nonvanishing result.
For a single insertion of the Lagrangian term
\be
\LRRR\ =\ c_{R^3}(\mu) \sqrt{-g}\, R^{\alpha \beta}{\!}_{\gamma\delta} 
R^{\gamma \delta}{\!}_{\rho\sigma} R^{\rho\sigma}{\!}_{\alpha\beta} \,,
\label{LcR3}
\ee
the identical-helicity matrix element is~\cite{vanNWu}
\be
M_4({+}{+}{+}{+})\ =\ - 60 \, i \, c_{R^3}(\mu)
 \, \left(\frac{\kappa}{2}\right)^4 \, \T^2 \, s_{12} s_{23} s_{13} \,,
\label{allplusfromR3}
\ee
where
\be
\T = \frac{[12] [34]} {\langle 1 2 \rangle \langle 3 4 \rangle} \,,
\label{TDef}
\ee
and $s_{12} = (k_1 + k_2)^2$, $s_{23} = (k_2 + k_3)^2$ and $s_{13} = (k_1 + k_3)^2$
are the usual Mandelstam invariants.  The factor $\T$ is a pure phase
constructed from the spinor products $\langle a b \rangle$ and $[a b]$,
defined in e.g.~Ref.~\cite{Review}.

Although no mass regulator was used in Ref.~\cite{PreviousPaper},
the Gauss--Bonnet operator (\ref{GaussBonnetDivergence})
contributes nonvanishing subdivergences, because internal legs of
the two-loop amplitude propagate in $D$ dimensions.
\Fig{CounterSingleDoubleFigure} illustrates the complete set of
counterterm contributions required to renormalize the
dimensionally-regulated four-graviton amplitude at two loops.  Besides the
bare amplitude in \Fig{CounterSingleDoubleFigure}(a),
there is the single insertion of the GB operator into a one-loop amplitude
in \Fig{CounterSingleDoubleFigure}(b)
and the double-GB-counterterm insertion into a tree amplitude,
\Fig{CounterSingleDoubleFigure}(c).
Finally, the two-loop $R^3$ counterterm insertion is shown in 
\Fig{CounterSingleDoubleFigure}(d).
All contributions shown are representative ones, out of a much larger number
of Feynman diagrams; for example, the bare contribution also includes
nonplanar diagrams.

For pure gravity, assembling the contributions from
\Fig{CounterSingleDoubleFigure}(a)--(c), the divergence in the
two-loop four-graviton amplitude 
and associated renormalization-scale dependence is~\cite{PreviousPaper}
\be
\M^{\twoloop}_4({+}{+}{+}{+})\Big|_{\rm (a)\hbox{--}(c)}
\ =\ \Bigl(\frac{\kappa}{2} \Bigr)^6 \frac{i}{(4 \pi)^4}
s_{12} s_{23} s_{13}  \T^2 \, \biggl(\frac{209}{24\, \eps} 
- \frac{1}{4}\, \ln\mu^2 \biggr)  + \hbox{finite}\,.
\label{DivMu}
\ee
In a minimal subtraction prescription, the effect of the
$R^3$ counterterm in \Fig{CounterSingleDoubleFigure}(d)
is simply to remove the $209/24 \times 1/\e$ term. 
Including matter fields, the ultraviolet divergence
changes under duality transformations~\cite{PreviousPaper}.
This change might not be surprising, given that
the coefficient of the one-loop Gauss--Bonnet
subdivergence~(\ref{GaussBonnetDivergence}) is not
invariant under duality transformations~\cite{DuffInequivalence}.
For example, adding $n_3$ three forms, which do not propagate in
four dimensions, changes the coefficient of the infinity in
\eqn{GSDiv} to
\be
\frac {209}{1440 \eps} \ \rightarrow \ 
\frac{209}{1440 \eps} - \frac{1}{8\eps}\, n_3 \,,
\label{DivergenceShift}
\ee
while the coefficient of $\ln\mu^2$ is unaltered.
Also, the value of the leading infinity depends nontrivially on the details of
the regularization procedure, while the coefficient of the $\ln\mu^2$ term
does not.

\begin{figure}[tb]
\captionsetup[subfigure]{labelformat=empty}
\begin{center}
\subfloat[\large (a)]{{\includegraphics[scale=.38]{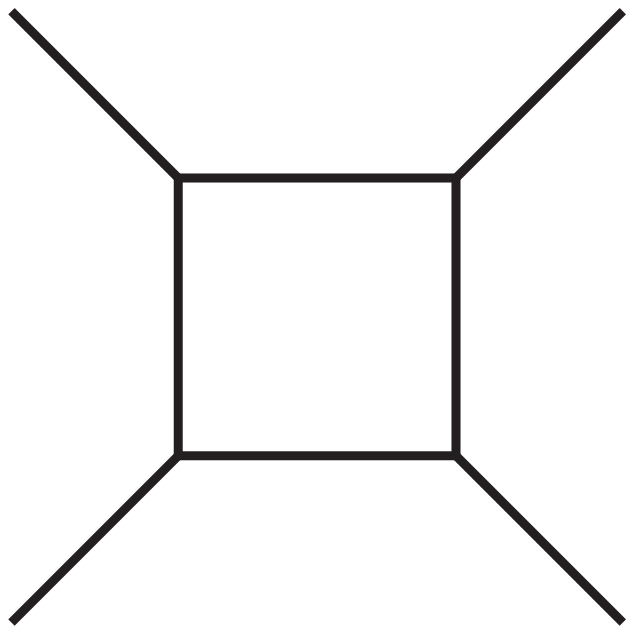}}}
\hspace{.5cm}
\subfloat[\large (b)]{\includegraphics[scale=.38]{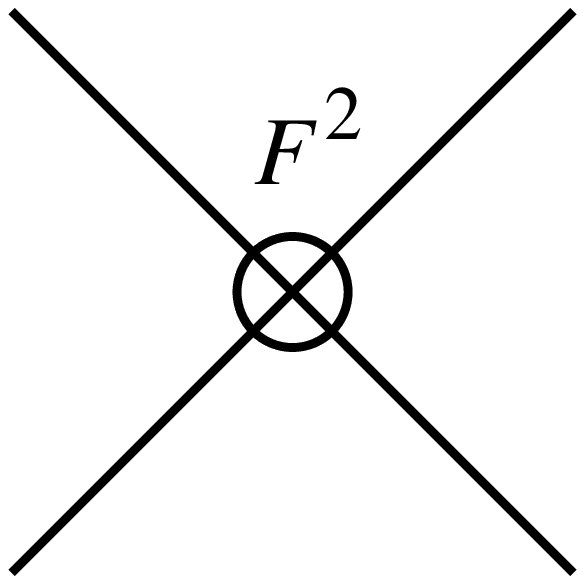}}
\end{center}
\vskip -.7cm 
\caption[a]{\small Renormalization of on-shell Yang--Mills 
amplitudes at one loop requires (a) the bare amplitude and 
(b) an $F^2$ counterterm, for which representative contributions are shown.
\label{YMFigure}
}
\end{figure}

The fact that the two numerical coefficients in \eqn{DivMu} are rather
different, and that one changes under duality transformations but not
the other, implies that they are not directly linked.  This is rather
curious.  From the textbook computation of the one-loop beta function
in Yang--Mills theory, we are used to the idea that they are linked.
In that case, the analog of \Fig{CounterSingleDoubleFigure} is
\Fig{YMFigure}.  To renormalize the on-shell amplitudes in the theory
at one loop, we need the bare one-loop amplitude, with a
representative diagram shown in \Fig{YMFigure}(a), and a single
insertion of the $F^a_{\mu\nu}F^{a\,\mu\nu}$ counterterm into a tree-level
amplitude, with a representative diagram shown in \Fig{YMFigure}(b).

Schematically, these two contributions depend on the
renormalization scale $\mu$ as follows:
\be
\frac{C^{\rm (a)}}{\eps} \, (\mu^2)^{\eps} 
 + \frac{C^{\rm (b)}}{\eps}
\ =\ \Big(C^{\rm (a)} + C^{\rm (b)} \Bigr) \frac{1}{\eps} + 
   C^{\rm (a)} \ln\mu^2 + \cdots,
\label{YMDivergence}
\ee
where the $(\mu^2)^{\eps}$ factor in the bare amplitude
compensates for the dimension of the loop integration measure
$d^{4-2\e}\ell$, where $\ell$ is the loop momentum.
In a minimal subtraction scheme, one chooses $C^{\rm (b)} = - C^{\rm (a)}$
to cancel the $1/\e$ pole.
Because the counterterm insertion has no factor of $(\mu^2)^{\eps}$,
the leading divergence $C^{\rm (a)}$ is tied directly to
the renormalization-scale dependence of the coupling, i.e.~the beta function,
independent of the details of the regularization procedure.  

What about gravity at two loops?
As explained in Ref.~\cite{PreviousPaper}, the disconnect between the
divergences and the renormalization-scale dependence happens
because of an interplay between the bare terms and the evanescent
subdivergences.  The analog of \eqn{YMDivergence} for the divergence
and $\ln\mu^2$ dependence of the two-loop gravity amplitude is
\be
\frac{C^{\rm (a)}}{\eps} \,(\mu^2)^{2 \eps} 
+ \frac{C^{\rm (b)}}{\eps}\, (\mu^2)^{\eps} 
 +  \frac{C^{\rm (c)}}{\eps} +  \frac{C^{\rm (d)}}{\eps} 
 = \Big(C^{\rm (a)} + C^{\rm (b)} + C^{\rm (c)} + C^{\rm (d)}  \Bigr) \frac{1}{\eps} + 
 (2 C^{\rm (a)} + C^{\rm (b)}) \ln\mu^2 + \cdots.
\label{Divergence}
\ee
The differing powers of $\mu$ for each contribution follow from
dimensional analysis of the integrals, after accounting for the fact
that the counterterm insertions do not carry factors of
$(\mu^2)^{\eps}$.

The coefficient of the $R^3$ counterterm $C^{\rm (d)}$
cancels the two-loop divergence in \eqn{Divergence}, as a consequence of the
renormalization conditions,
$C^{\rm (d)} = - C^{\rm (a)} - C^{\rm (b)} - C^{\rm (c)}$.
In the amplitude computed in
Ref.~\cite{PreviousPaper}, the value of the coefficient of the
two-loop $R^3$ counterterm depends on duality transformations, while
the coefficient in front of the $\ln\mu^2$, namely $2 C^{\rm (a)} + C^{\rm (b)}$,
does not.  The fact that
different combinations of coefficients appear in the divergence and
in the $\ln \mu^2$ term explains why the two-loop divergence and
renormalization-scale dependence do not have to be simply related.
As we discuss in the next section, the coefficient of the logarithm
can be computed directly in four dimensions,
completely avoiding the issue of evanescent operators.
On the other hand, the divergence is exposed to the
subtleties of evanescent operators and dimensional regularization.
More remarkably, as found in a variety of examples~\cite{PreviousPaper},
the $\ln\mu^2$ coefficient satisfies a simple
formula, which we explain in the next section.

The disconnect between the divergence and the renormalization-scale dependence
could lead to situations where an explicit divergence is present, yet there
is no associated running of a coupling or other physical
consequences. As an example, we have computed the divergence in
${\cal N}=1$ supergravity with one matter multiplet using the same
techniques.  It is convenient to include a matter multiplet because
for this theory we can construct the two-loop integrand
straightforwardly using double-copy techniques~\cite{BCJ}.
Even though this theory is supersymmetric, the trace anomaly is
nonvanishing~\cite{DuffWeyl}.  Therefore there are subdivergences
of the form of \Fig{CounterSingleDoubleFigure}(b),
as well as \Fig{CounterSingleDoubleFigure}(c).
We have computed the four contributions corresponding to
\fig{CounterSingleDoubleFigure}.  They are given by
\be
C^{\rm (a)} = \frac{11}{16}\,, \hskip 1 cm
C^{\rm (b)} = -\frac{11}{8}\,, \hskip 1 cm
C^{\rm (c)} = \frac{363}{32} \,, \hskip 1 cm
C^{\rm (d)} = -\frac{341}{32} \,,
\label{Neq1numbers}
\ee
where the normalization corresponds to
$C^{\rm (a)} + C^{\rm (b)} + C^{\rm (c)} = 209/24$ for pure gravity;
see \eqn{DivMu}.
So the divergence from terms (a)--(c) in \eqn{Divergence} is nonzero,
but the $\ln\mu^2$ coefficient vanishes, $2C^{\rm (a)} + C^{\rm (b)} = 0$.
In fact, it turns out that all logarithms $\ln s_{ij}$ in the amplitude
cancel as well.  The polynomial terms can be canceled by the same $R^3$
counterterm but with a finite coefficient (or equivalently,
an order $\e$ correction to $C^{\rm (d)}$).

The upshot is that for this ${\cal N}=1$ supergravity theory, the
divergence and associated trace anomaly has the curious effect of
violating the supersymmetry Ward identity~\cite{SWI} that requires the
identical-helicity amplitude to vanish. The appearance of a
divergence is due to the breaking of supersymmetry by the trace
anomaly, which induces subdivergences even when supersymmetry implies
that no divergences can be present~\cite{TwoLoopSusy}.  To restore the
supersymmetry Ward identities requires adding an $R^3$ counterterm to
the theory, with both a $1/\e$ and a finite coefficient, which fixes
the two-loop amplitude uniquely.  This procedure is possible only
because the $\ln\mu^2$ coefficient vanishes.  That is, in this case
there is no loss of predictivity, even though there is a $1/\e$
divergence.  If the $\ln\mu^2$ coefficient is nonvanishing, as in the
case of pure gravity, then there must be an arbitrary finite constant
in the renormalization procedure, associated with fixing the $R^3$ coupling at
different choices of renormalization scale, leading to the usual loss
of predictivity of nonrenormalizable theories.

This discussion applies more generally. Suppose there is a hidden
symmetry that would enforces finiteness if it can be preserved.  Yet
if that symmetry is broken by the trace anomaly, or more generally by
the regularization procedure, we might conclude that the theory's
divergence implies a loss of predictivity.  It is therefore always
crucial to inspect the renormalization-scale dependence.

In contrast, one might even imagine a
regularization prescription that eliminates the $1/\e$ divergence, 
for example by making the perverse
choice $n_3 = 8 \cdot 209/1440$ in \eqn{DivergenceShift} for
the case of pure gravity. However, since the $\ln\mu^2$ coefficient is
nonvanishing in this case, there is still an arbitrariness in
the finite $R^3$ counterterm associated with different choices for $\mu$,
and an associated loss of predictivity.  The theory is no better 
than an ultraviolet-divergent theory, even if the $1/\e$ divergence
is arranged to cancel.

From now on we focus entirely on the renormalization-scale dependence.
For the two-loop graviton identical-helicity scattering amplitude
with various matter content, Ref.~\cite{PreviousPaper} found the
following simple form:
\be
\M^{\twoloop}_4({+}{+}{+}{+}) \Bigr|_{\ln\mu^2}
\ =\ - \Bigl(\frac{\kappa}{2} \Bigr)^6 \frac{i}{(4 \pi)^4}
s_{12} s_{23} s_{13}  \T^2 \, \frac{N_b-N_{\! f}}{8} \, \ln\mu^2\,,
\label{NbNf}
\ee
where $N_b$ and $N_{\! f}$ are the number of physical
four-dimensional bosonic and fermionic states in the theory.
Using \eqn{allplusfromR3}, this result is equivalent to the
running of the $R^3$ coefficient according to
\be
\mu \frac{\partial \, c_{R^3}}{\partial\mu} 
\ =\ \Bigl(\frac{\kappa}{2} \Bigr)^2 \frac{1}{(4 \pi)^4}
\, \frac{N_b-N_{\! f}}{240} \,.
\label{NbNfcR3}
\ee
Because the number of physical four-dimensional states does not
change under duality transformations, this equation
is automatically independent
of such transformations and of the details of the regularization scheme.
In fact, the result was only confirmed in Ref.~\cite{PreviousPaper}
for minimally-coupled scalars, antisymmetric tensors and (non-propagating)
three-form fields.  The generalization to fermionic contributions
was based on the previously-mentioned supersymmetry Ward identities.
It is quite remarkable that such a simple formula for the
renormalization-scale dependence emerges from the computations 
carried out in Ref.~\cite{PreviousPaper}.
How did this happen?  We answer this in the next section.


\section{Renormalization-scale dependence directly from four-dimensional unitarity cuts}
\label{FourDimSection}

In this section we explain the simple form of the
renormalization-scale dependence in \eqn{NbNf} using four-dimensional
unitarity cuts.  We show that it holds for any massless theory with
minimal couplings to gravity and renormalizable matter interactions.
From simple dimensional considerations, contributions to the $R^3$
operator necessarily involve couplings with the dimension of the
gravitational coupling $\kappa$, which carries the dimension of
inverse mass, $1/M_P$.  Renormalizable matter interactions are either
dimensionless or carry the dimension of mass, so they can contribute
only to lower-dimension operators than $R^3$ at two loops, and
therefore they are not relevant at this order.  We will also explain
why dilatons and antisymmetric tensors---whose minimal couplings to
gravitons have two derivatives, as does pure gravity---also respect
\eqn{NbNf}, as found in the computations of
Refs.~\cite{PreviousPaper,DoubleCopyGravity}.

Unitarity cuts are not directly sensitive to the $\ln\mu^2$
dependence.  However, in a massless theory, simple dimensional analysis
relates the coefficient of $\ln\mu^2$ to the coefficients of
logarithms of kinematic invariants, $\ln s_{ij}$,
because the arguments of all logarithms need to be dimensionless.
Because the coefficient of $\ln\mu^2$ is finite, we can evaluate the
unitarity cuts in four dimensions (after subtracting a universal
infrared divergence).  Thus we automatically avoid evanescent operators,
such as the Gauss--Bonnet term (\ref{GaussBonnetDivergence}).  Our approach
greatly clarifies the essential physics, showing that duality transformations
cannot change the logarithms in the scattering amplitude, because in
four dimensions, unlike $D$ dimensions, duality does not change the Lorentz
properties or number of physical states.
The calculation of the logarithms using unitarity cuts was carried out
long ago by Dunbar and Norridge~\cite{DunbarNorridge}.  Recently a
similar technique has been applied to two-loop identical-helicity
amplitudes in gauge theory by Dunbar, Jehu and
Perkins~\cite{DunbarJehuPerkins}.
Here we repeat the
two-loop four-graviton calculation, but in a way that completely avoids
dimensional regularization and focuses on the consequences and
interpretation of the renormalization scale.

\begin{figure}[tb]
\captionsetup[subfigure]{labelformat=empty}
\begin{center}
\subfloat[\large(a)]{\includegraphics[scale=.48]{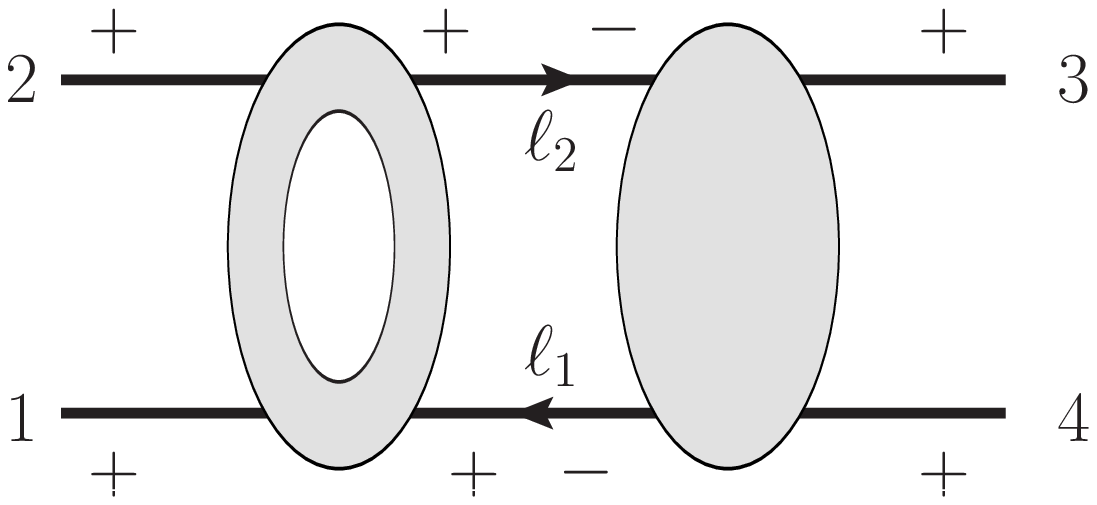}}
\hspace{.3cm}
\subfloat[\large(b)]{\includegraphics[scale=.48]{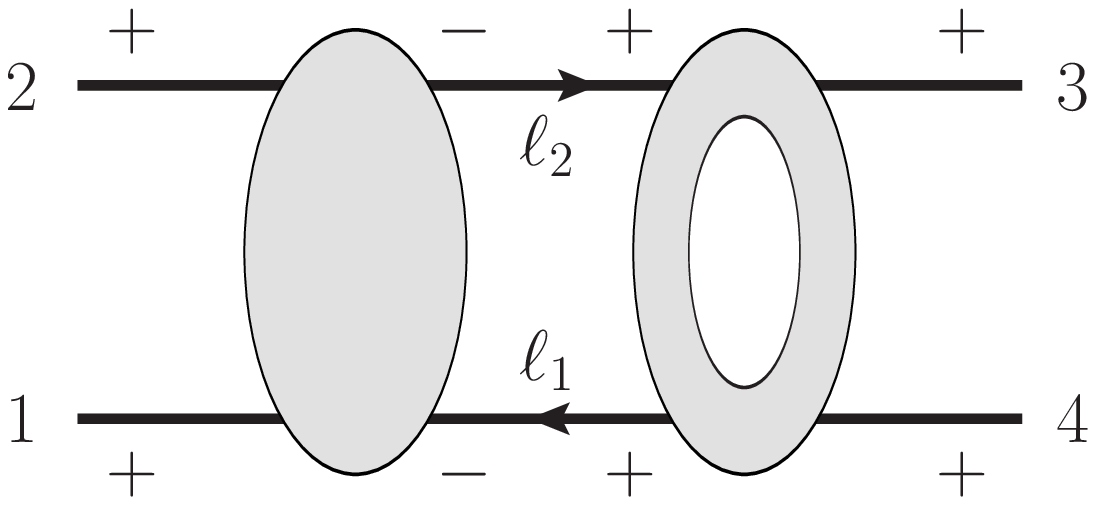}}
\end{center}
\vskip -.7cm 
\caption{The $s$-channel two-particle cuts (a) and (b) from which we
  can extract the logarithmic parts of the two-loop four-point
  identical-helicity four-graviton amplitude. The exposed lines are
  placed on shell and are in four dimensions.}
\label{TwoCutFigure}
\end{figure}

\begin{figure}[tb]
\includegraphics[scale=.5]{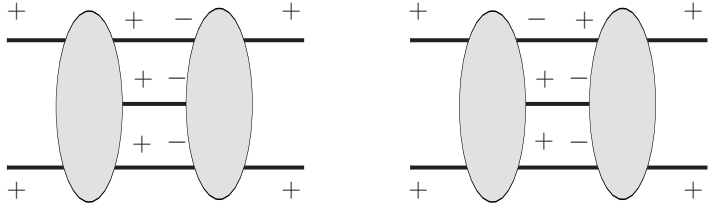}
\caption{Representative contributions to the three-particle cut. This
  cut generates no new $\ln \mu^2$ contributions to the $R^3$
  operator for the identical-helicity four-graviton amplitude.}
\label{ThreeCutFigure}
\end{figure}

We obtain the kinematical logarithms of the all-plus helicity amplitude
from the four-dimensional unitarity cuts.  At two loops, there
are cuts where two particles cross the cut, illustrated in
\fig{TwoCutFigure}, and where three particles cross the cut, shown in
\fig{ThreeCutFigure}.  In four dimensions, many contributions to these cuts
vanish, because the tree amplitude on one side of a cut vanishes. 

In pure gravity, all contributions to the three-particle cuts
shown in \fig{ThreeCutFigure} vanish, because they contain a either
a tree amplitude with all identical helicities, or one with one leg of
opposite helicity.  Such five-graviton tree amplitudes vanish.
Adding minimally-coupled matter does not alter this conclusion.
As already noted, adding matter with renormalizable self
couplings cannot affect the coefficient of the $R^3$ operator.
Similarly, dilatons and antisymmetric tensors, with their minimal
couplings to each other and to gravity also cannot contribute,
because their amplitudes have similar vanishings as the pure gravity case,
where a pair of external (pseudo)scalar state should be assigned one
plus and one minus helicity.  All of these vanishings can be understood
from the fact that all such amplitudes can be constructed from
minimally-coupled gauge theory via the Kawai--Lewellen--Tye (KLT)
relations~\cite{KLT}, which all have the corresponding vanishings.
Alternatively, such tree amplitudes can be embedded into ${\cal N}=8$
supergravity, and then the supersymmetry Ward identities~\cite{SWI}
imply the required vanishings.

The two-particle cut does have nonvanishing contributions; however, the
cut lines have to be gravitons, with the helicity configurations
displayed in \fig{TwoCutFigure}.  If a massless particle other than a graviton
crosses the cut with this helicity configuration, then the tree
amplitude entering the cut necessarily vanishes.  These vanishings can be
understood in various ways. The KLT decomposition offers one such way.
Consider the KLT decomposition of the gravitational tree amplitude
on the right-hand side of \fig{TwoCutFigure}(a) into a
product of two gauge-theory amplitudes~\cite{KLT},
\be
M^\tree(\ell_1,-\ell_2,3,4) = s_{12}\, A^\tree(\ell_1,-\ell_2,3,4)
 A^\tree(\ell_1,-\ell_2,4,3)\,,
\ee
where $M^\tree(1,2,3,4)$ is the gravitational tree amplitude and
$A^\tree(1,2,3,4)$ is a color-ordered Yang--Mills tree amplitude.  (In
this expression the couplings are stripped off.)  If legs $3$ and $4$
of the gravitational amplitude are positive-helicity gravitons in an
all-outgoing convention, then the corresponding legs in the
gauge-theory amplitudes are positive-helicity gluons, so that the spins
match.  For gauge-theory amplitudes where legs $3$ and $4$ are
positive-helicity gluons, the only nonvanishing configuration is
where the remaining two legs are negative-helicity gluons.  The KLT
relations then imply that the only nonvanishing gravity tree amplitude
is when the two legs labeled by $\ell_1$ and $-\ell_2$ in the unitarity cut
are gravitons with negative helicity.
Other configurations, corresponding to particles other than
negative-helicity gravitons, vanish because at least one of the corresponding
gauge-theory amplitudes vanishes.

A consequence of these
restrictions is that the one-loop amplitude appearing on the other
side of the two-particle cut must be an all-plus-helicity amplitude
with only external gravitons. Such amplitudes are remarkably
simple~\cite{DunbarNorridge}.  This simplicity enormously streamlines
the calculation of the cut.  There are two contributions to the
$s_{12}$-channel cut, shown in \fig{TwoCutFigure}(a) and (b),
depending on whether the loop amplitude is located on the left or
right side of the cut.  However, they give equal contributions,
because \fig{TwoCutFigure}(b) can be mapped back to
\fig{TwoCutFigure}(a) by relabeling the momenta by $k_i \to k_{i+2}$,
where the indices are modulo 4, and we will see that the cut is
invariant under this operation.  In addition to the $s_{12}$-channel
cut displayed in \fig{TwoCutFigure}, there are also cuts in the $s_{23}$ and
$s_{13}$ channels, which can be obtained from the $s_{12}$ channel by
Bose symmetry, permuting $k_1 \lr k_3$ and $k_1 \lr k_4$,
respectively.

The required one-loop amplitude with four identical-helicity gravitons
is~\cite{DunbarNorridge},
\be
M^{\oneloop}(1^+,2^+,3^+,4^+)\ =\  
-\frac{i}{(4 \pi)^{2}}  \frac{N_{b} - N_{\!f} }{240} 
        \biggl(\frac{\kappa}{2}\biggr)^4\,
      {\cal T}^2 \, (s_{12}^2 + s_{14}^2 + s_{24}^2) \,,
\label{gravity1loop} 
\ee
where the permutation-invariant, pure-phase spinor combination $\T$
is defined in \eqn{TDef}.  The one-loop external graviton
amplitude is unaffected by any interactions of the matter fields in a
minimally-coupled theory: at one loop with all external gravitons
there are no diagrams containing matter self-interactions.

In Yang--Mills theory, Bardeen and Cangemi~\cite{AllPlusAnomaly}
argued that the corresponding identical-helicity amplitude is
nonvanishing because of an anomaly in the infinite-dimensional
symmetry of the self-dual sector of the theory.  Presumably, the same
holds in gravity.  It is quite interesting that this anomaly-like
behavior appears crucial for obtaining a nonvanishing one-loop
four-graviton amplitude, which as we will see below leads to a nonvanishing
coefficient of the $\ln\mu^2$ term.

We also need the four-graviton tree amplitude.  It is easily
obtained from the KLT relation~\cite{KLT},
\bea
M^{\tree}(1^-,2^-,3^+,4^+) &=&  -i \biggl(\frac{\kappa}{2}\biggr)^2
 s_{12} \, A^\tree(1,2,3,4) A^\tree(1,2,4,3)
 \nonumber  \\
&=& i \biggl(\frac{\kappa}{2}\biggr)^2 s_{12} \,
  \frac{{\spa1.2}^3}{\spa2.3 \spa3.4 \spa 4.1} \,
   \frac{{\spa1.2}^3}{\spa2.4\spa4.3\spa3.1}\,.
\label{gravitytree}
\eea

We now calculate the unitarity cut in \fig{TwoCutFigure}(a).  The cut
integrand is given by the relabeled product of
\eqns{gravity1loop}{gravitytree},
\bea
C_{12} &=& {\cal N} \, s_{12}\, (s_{12}^2 + s_{1\ell_1}^2 + s_{2\ell_1}^2) 
\nonumber \\
&& \times
\biggl(\frac{\spb1.2 \spb{\ell_2}.{(-\ell_1)}}
            {\spa1.2 \spa{\ell_2}.{(-\ell_1)}} \biggr)^2 
\frac{\spa{\ell_1}.{(-\ell_2)}^3}{\spa{(-\ell_2)}.3 \spa3.4 \spa4.{\ell_1}} \, 
\frac{\spa{\ell_1}.{(-\ell_2)}^3}{\spa{(-\ell_2)}.4 \spa4.3 \spa3.{\ell_1}}\,,
\eea
where the labels follow \fig{TwoCutFigure}(a)
and the normalization factor is
\be
{\cal N} = \frac{1}{(4\pi)^2}  \frac{N_b - N_{\! f}}{240}
  \biggl(\frac{\kappa}{2}\biggr)^6\,.
\ee
Rearranging the spinor products and using the identity 
$1/\spa{a}.{b} = [ba]/(k_a + k_b)^2$  gives
\bea
C_{12} &=&  {\cal N} \, \T^2 \, s_{12} \, 
 (s_{12}^2 + s_{1\ell_1}^2 + s_{2\ell_1}^2) \nonumber \\
&& \times 
\frac{\spa{\ell_1}.{\ell_2}  \spb{\ell_2}.3 \spa3.4 \spb4.{\ell_1}
      \spa{\ell_1}.{\ell_2} \spb{\ell_2}.4 \spa4.3\spb3.{\ell_1}} 
 {(\ell_2 - k_3)^2 (\ell_1 + k_4)^2 (\ell_2 -k_4)^2 (\ell_1 + k_3)^2 }\,.
\eea
The net effect of replacing $-\ell_1$ and $-\ell_2$ with $\ell_1$ and
$\ell_2$ is a factor of $+1$.  We can simplify $C_{12}$ further by
observing that the numerator forms a trace,
\bea
\spa{\ell_1}.{\ell_2}\spb{\ell_2}.3 \spa3.4 \spb4.{\ell_1}\spa{\ell_1}.{\ell_2} 
   \spb{\ell_2}.4 \spa4.3\spb3.{\ell_1}
&=& \frac{1}{2} \tr[(1-\gamma_5)\ell_1 \ell_2 k_3 k_4 \ell_1 \ell_2 k_4 k_3]
 \nonumber\\
&=& (\ell_1 +k_3)^2 (\ell_1 + k_4)^2  s_{34}^2 \,,
\eea
where we used $\ell_2 = \ell_1 + k_3 + k_4$ and the on-shell
conditions $\ell_1^2 = \ell_2^2 = 0$ to simplify the trace. 
Thus, the numerator cancels the (doubled) propagators leaving
\bea
C_{12} &=&  {\cal N}  \T^2 
  \, s_{12}^3 \, \frac{s_{12}^2 + s_{1\ell_1}^2 + s_{2\ell_1}^2} 
     {(\ell_2 - k_3)^2 (\ell_2 -k_4)^2 } \nonumber\\
 & = & - {\cal N} \T^2 \, s_{12}^2  
 \, (s_{12}^2 + s_{1\ell_1}^2 + s_{2\ell_1}^2)
 \biggl[ \frac{1}{(\ell_1 + k_4)^2} +  \frac{1}{(\ell_1 + k_3)^2}
        \biggr] \,.
\label{Cut12}
\eea

This expression for the cut actually has an infrared divergence when
integrated over phase space.  However, this divergence is harmless because
infrared singularities of gravity theories are relatively
simple~\cite{IRPapers}.  The source of the singularity is from
exchange of soft virtual gravitons with momentum $\ell_1+k_3$ or $\ell_1+k_4$;
the soft limit is when $\ell_1 \rightarrow -k_4$ or
$\ell_1 \rightarrow -k_3$, for the first or second term in \eqn{Cut12},
respectively.
To remove the infrared singularity, we simply subtract the
soft limit of the integrand, replacing $C_{12}$ by
\be
\tilde C_{12} = - {\cal N} \T^2 \, s_{12}^2
 \, \frac{ s_{1\ell_1}^2 + s_{2\ell_1}^2 -  s_{14}^2 - s_{24}^2 }
       { (\ell_1 + k_4)^2} 
 + (k_3 \lr k_4).
\ee
The subtraction terms correspond to cut scalar triangle integrals.
Since the triangle integrals that are subtracted converge in the
ultraviolet, the subtraction has no effect on the ultraviolet
logarithms with which we are concerned here.

The discontinuity is obtained by integrating over the Lorentz-invariant
phase space,
\be
I_{12}\ =\ \int {\rm dLIPS} \, \tilde C_{12}
\ =\ - {\cal N} \T^2 \, s_{12}^3 \hat{I}_{12}
 + (k_3 \lr k_4)\,,
\label{introducehatI12}
\ee
where
\be
\hat{I}_{12}\ =\ \int {\rm dLIPS} \, 
  \frac{ (2k_1\cdot\ell_1)^2 + (2k_2\cdot\ell_1)^2 -  s_{14}^2 - s_{24}^2 }
       { s_{12} \, (2k_4\cdot\ell_1) } \,.
\label{hatI12}
\ee
We perform the phase-space integration in the center-of-mass frame,
parametrizing the external momenta as
\bea
k_1&=&\frac{\sqrt{s}}{2}(-1,\,\sin\theta\cos\phi,
        \,\sin\theta\sin\phi,\,\cos\theta)\,, \nonumber \\
k_2&=&\frac{\sqrt{s}}{2}(-1,\,-\sin\theta\cos\phi,
        \,-\sin\theta\sin\phi,\,-\cos\theta)\,, \nonumber \\
k_3&=&\frac{\sqrt{s}}{2}(1,\,0,\,0,\,1)\,, \nonumber \\
k_4&=&\frac{\sqrt{s}}{2}(1,\,0,\,0,\,-1)\,,
\label{kpara}
\eea
and the internal momentum as
\bea
\ell_1&=&\frac{\sqrt{s}}{2}(-1,\,\sin\hat\theta\cos\hat\phi,
        \,\sin\hat\theta\sin\hat\phi,\,\cos\hat\theta)\,,
\eea
while $-\ell_2^0=\ell_1^0$ and $-\vec\ell_2=-\vec\ell_1$.
The on-shell conditions enforce the constraints
$|\ell_i^0| = |\vec\ell_i| = \sqrt{s}/2$, $i=1,2$.
The standard two-body phase-space measure is
\be
\int {\rm dLIPS} 
\ =\ \frac{1}{2}\frac{1}{8\pi}\int_{-1}^{1} \frac{d\cos\hat\theta}{2}
   \int_0^{2\pi}\frac{d\phi}{2\pi} \,.
\ee 
There is an extra Bose symmetry factor of $1/2$ because two
identical-helicity gravitons cross the cut.
Substituting the momentum parametrization into \eqn{hatI12}
gives an expression for $\hat{I}_{12}$ purely in terms of angular variables,
which can be integrated easily,
\bea
 \hat{I}_{12} &=& \frac{1}{16\pi}
\int_{-1}^1 \frac{d\cos\hat\theta}{2} \int_0^{2\pi} \frac{d\hat\phi}{2\pi}
\frac{1}{1-\cos\hat\theta}
\Bigl[ \cos^2\theta\,\sin^2\hat\theta
 - \sin^2\theta\,\sin^2\hat\theta\,\cos^2(\phi-\hat\phi)\nonumber\\
&&\hskip5.5cm\null
- \frac{1}{2} \sin2\theta\,\sin2\hat\theta\,\cos(\phi-\hat\phi)
\Bigr] \nonumber\\
&=& \frac{1}{16\pi}
\int_{-1}^1 \frac{d\cos\hat\theta}{2}
\frac{1}{1-\cos\hat\theta}
\Bigl[ \cos^2\theta\,\sin^2\hat\theta
 - \frac{1}{2} \sin^2\theta\,\sin^2\hat\theta \Bigr] \nonumber\\
&=& \frac{1}{16\pi}
\Bigl[ \cos^2\theta - \frac{1}{2} \sin^2\theta \Bigr]
\int_{-1}^1 \frac{d\cos\hat\theta}{2} [ 1 + \cos\hat\theta ] \nonumber\\
&=& \frac{2 - 3 \sin^2\theta}{32\pi} \,.
\eea
Using $s_{13}s_{23} = (s_{12}^2/4)\times \sin^2\theta$, we can re-express the 
answer in a Lorentz-invariant form:
\be
 \hat{I}_{12}\ =\ \frac{1}{16\pi} \frac{s_{12}^2 - 6s_{13}\,s_{23}}{s_{12}^2} \,.
\ee
Since this result is invariant under $k_3 \rightarrow k_4$,
the exchange contribution in \eqn{introducehatI12} just gives a factor of 2.

Putting it all together, we have
\bea
  \tilde{C}_{12}
 &=& - \frac{\N \T^2}{8\pi} \ s_{12} \, (s_{12}^2 - 6 s_{13}\,s_{23})\\
  &=& 2\pi i \left[ \frac{i}{(4 \pi)^4} \frac{N_b - N_{\!f}}{240}
  \left(\frac{\kappa}{2}\right)^6 \T^2 s_{12}(s_{12}^2 - 6 s_{13}\, s_{23})\right]
  \label{IPiResult}.
\eea
We extracted a factor of $2\pi i$ because the analytic continuation of
$\ln(-s_{ij}/\mu^2)$ from below the cut ($s_{ij} \to s_{ij} - i\varepsilon$)
to above the cut ($s_{ij} \to s_{ij} + i\varepsilon$) is
\be\label{log}
\ln\biggl(\frac{-s_{ij}}{\mu^2}\biggr) 
\ \to\ \ln\biggl(\frac{-s_{ij}}{\mu^2}\biggr) - 2\pi i\,.
\ee
Thus, the $s_{12}$-channel discontinuity we computed is related to the coefficient
of $\ln\mu^2$ by
\be
M^{\rm 2-loop}|_{\ln\mu^2}
\ =\ \frac{1}{2\pi i} M^{\rm 2-loop}|_{\rm disc} \times \ln \mu^2\,.
\label{discrule}
\ee
We also need to multiply by a factor of 2 for the contribution of
\fig{TwoCutFigure}(b), and include the contributions of the other two
channels, using
\be
s_{12}(s_{12}^2 - 6 s_{13}\, s_{23})\ +\ (k_1\lr k_3)\ +\ (k_1\lr k_4)
= s_{12}^3 + s_{23}^3 + s_{13}^3 - 18 s_{12} s_{23} s_{13}
= - 15 s_{12} s_{23} s_{13} \,.
\ee
We obtain 
\be
M^{\rm 2-loop}({+}{+}{+}{+})\Bigl|_{\ln\mu^2}
\ =\ - \left(\frac{\kappa}{2}\right)^6
\frac{i}{(4 \pi)^4} s_{12} s_{23} s_{13} \T^2 \, \frac{N_b-N_f}{8} \ln\mu^2.
\label{FinalMu}
\ee

Thus, we have derived the simple renormalization-scale dependence of
the two-loop four-graviton amplitude~\cite{PreviousPaper},
but now in a way that avoids reliance on
evanescent operators or other subtleties of dimensional
regularization.  Given that only four-dimensional quantities were
used, duality transformations manifestly
cannot affect the renormalization-scale dependence.

\section{Conclusions}
\label{ConclusionSection}

In this paper we explained the simple form of the
renormalization-scale dependence of two-loop gravity amplitudes
proposed in Ref.~\cite{PreviousPaper}.  While the two-loop ultraviolet
divergence in dimensional regularization changes under duality
transformations, and is afflicted by evanescent subdivergences, the
renormalization-scale dependence is remarkably
simple~\cite{PreviousPaper}.  In order to explain its simple form, we
used four-dimensional unitarity cuts, which effectively converted the
two-loop computation into a one-loop one.  As in
Ref.~\cite{PreviousPaper}, we studied the identical-helicity
amplitude, because it is particularly simple to evaluate, yet is
sensitive to the two-loop $R^3$ ultraviolet divergence.  While the
renormalization scale $\ln\mu^2$ does not itself have a unitarity cut,
on dimensional grounds its coefficient must balance the coefficients
of the logarithms of kinematic variables, thus allowing us to extract
the $\ln\mu^2$ coefficient directly from the unitarity cuts.  This
method avoids the need for ultraviolet regularization, as well as all
subtleties associated with evanescent operators.  A trivial integral
over the two-body phase space for intermediate gravitons is all that
is required to explain the simple formula (\ref{FinalMu}) 
of Ref.~\cite{PreviousPaper}.

A rather interesting property of the gravity divergence is that it
appears to be tied to an anomaly.  In Yang--Mills theory, the
nonvanishing of the one-loop identical helicity amplitude has been
tied to an anomaly in the conserved currents of self-dual Yang--Mills
theory~\cite{AllPlusAnomaly}.  We expect gravity to be similar.
Integrability has been used to construct classical self-dual solutions
to Einstein's equations~\cite{SDGIntegrable}.  It is natural to
conjecture that a quantum anomaly in the conservation of the
associated currents of self-dual gravity~\cite{Krasnov} could be
responsible for the non-vanishing one-loop
amplitude~(\ref{gravity1loop}) which underlies the two-loop $\ln\mu^2$
dependence.  In any case, not only the two-loop divergence but the
nonvanishing of the one- and two-loop identical-helicity amplitudes
can be traced to an $\epsilon/\epsilon$ effect in dimensional
regularization, similar to the way that chiral and other anomalies
arise.  It would be quite enlightening if we could link the pure
gravity divergence, or more importantly, the nonvanishing
renormalization-scale dependence, more directly to an anomaly.

In this paper we considered the identical-helicity amplitude, because
it is the simplest helicity configuration that is sensitive to
the $R^3$ divergence.  It would be interesting to evaluate the other
helicity configurations to corroborate our understanding.  The other
helicity configurations are significantly more complicated, because the
three-particle cut no longer vanishes in four dimensions.  However,
the $({-}{+}{+}{+})$ helicity configuration, which also receives
contributions from the $R^3$ operator, should be tractable
using four-dimensional unitarity cuts.

Usually in field theory, the first dimensionally-regulated divergence
that is encountered is directly related to the renormalization-scale
dependence of either a coupling (i.e.~the beta function) or the coefficient
of an operator (i.e.~its anomalous dimension).  Pure Einstein gravity
at two loops provides an explicit counterexample to this expectation,
but it is probably not the only one.  As we discussed in
\sect{ReviewSection}, the key feature is that a candidate operator
for a first divergence is evanescent, vanishing in four
dimensions but not in $D$ dimensions.  The different $\mu$ dependence
associated with the bare and counterterm contributions spoils
the textbook relation between the pole in $\epsilon$ and the
renormalization-scale dependence at the following loop order.
Another place this might happen is in the effective field theory
of long flux tubes~\cite{OferZohar}.
The key lessons are that ultraviolet divergences in dimensional
regularization have to be treated with caution in certain circumstances,
and that it is safer to focus on the more physical renormalization-scale
dependence of the renormalized theory.

\subsection*{Acknowledgments}

We thank Clifford Chueng, Scott Davies, David Kosower and Josh Nohle for many useful
and interesting discussions.  We also thank Michael Duff for
pointing out the curious issue posed by a non-vanishing trace anomaly
in ${\cal N}=1$ supergravity.
This material is based upon work supported by the Department
of Energy under Award Number DE-{S}C0009937 and contract
DE-AC02-76SF00515.


\end{document}